\documentclass[nofootinbib,prd]{revtex4}%
\usepackage{amsmath}
\usepackage{amsfonts}
\usepackage{amssymb}
\usepackage{graphicx}%
\begin{document}
\title{Modified Dispersion Relations and Black Hole Entropy}
\author{Remo Garattini}
\email{Remo.Garattini@unibg.it}
\affiliation{Universit\`{a} degli Studi di Bergamo, Facolt\`{a} di Ingegneria, Viale
Marconi 5, 24044 Dalmine (Bergamo) ITALY.}
\affiliation{INFN - sezione di Milano, Via Celoria 16, Milan, Italy.}

\begin{abstract}
We compute the black hole entropy in the context of the Modified Dispersion
Relations using the brick wall model. An explicit dependence of the radial
coordinate approaching the horizon is shown to analyze the behavior of the
divergence. We find that, due to the modification of the density of states,
the brick wall can be eliminated. By assuming a specific form for the radial
coordinate $r\left(  E/E_{P}\right)  $, we examine a possible candidate for
$r\left(  E/E_{P}\right)  $. A comparison with the 't Hooft approach is presented.

\end{abstract}
\maketitle

\section{Introduction}

After almost thirty years after the introduction of the famous
Bekenstein-Hawking formula\cite{Bekenstein,Hawking}%
\begin{equation}
S_{BH}=\frac{1}{4}A/l_{P}^{2},
\end{equation}
relating the entropy of a black hole and its area, the thermodynamics of such
objects still attracts research in this direction. One reason is due to the
lack of a Quantum Gravity theory which should be able to explain black hole
physics. Another reason comes from the fact that Hawking
radiation\cite{Hawking} develops modes of arbitrarily high frequency near the
horizon. The appearance of a trans-Planckian physics in Black Hole
thermodynamics has led many authors to consider that some deep change in
particle physics should come into play. In connection to this idea, in recent
years, there has been a proposal on how the fundamental aspects of special
relativity can be modified at very high energies. This modification has been
termed \textit{Doubly Special Relativity} (DSR)\cite{GAC}. In DSR, the Planck
mass is regarded as an observer independent energy scale. This assumption has
as effect in momentum space that the usual dispersion relation for a massive
particle of mass $m$ is changed into the following expression%
\begin{equation}
E^{2}g_{1}^{2}\left(  E/E_{P}\right)  -p^{2}g_{2}^{2}\left(  E/E_{P}\right)
=m^{2},\label{mdisp}%
\end{equation}
where $g_{1}\left(  E/E_{P}\right)  $ and $g_{2}\left(  E/E_{P}\right)  $ are
two functions which have the following property%
\begin{equation}
\lim_{E/E_{P}\rightarrow0}g_{1}\left(  E/E_{P}\right)  =1\qquad\text{and}%
\qquad\lim_{E/E_{P}\rightarrow0}g_{2}\left(  E/E_{P}\right)  =1.\label{prop}%
\end{equation}
The usual dispersion relation is recovered at low energies. Eqs.$\left(
\ref{mdisp},\ref{prop}\right)  $ are a representation of \textquotedblleft%
\textit{Modified Dispersion Relations\textquotedblright} (MDRs). The common
motivation in using them is in that they can be used as a phenomenological
approach to investigate physics at the Planck scale, where General Relativity
is no longer reliable. Concerning Black Hole thermodynamics, it has been
proved that the spectrum emitted at infinite distance from the hole is only
marginally affected by MDRs \cite{Unruh}. Nevertheless, when we consider the
statistical thermodynamics of quantum fields in the Hartle-Hawking state (i.e.
having the Hawking temperature $T_{H}$ at large radii), to keep under control
the high frequency divergences coming from the horizon sector, we need some
kind of cut-off of Planckian size known as \textquotedblleft\textit{brick
wall}\textquotedblright\cite{tHooft}. In a series of papers, it has been
suggested that this divergence could be absorbed in a renormalization of
Newton's constant\cite{SusUgl,BarEmp,EWin}, while other authors approached the
problem of the divergent brick wall using Pauli-Villars
regularization\cite{DLM,FS,KKSY}. Another interesting proposal comes from
non-commutative geometry which introduces a natural thickness of the horizon
replacing the 't Hooft's brick wall\cite{BaiYan}. Other successful attempts
come by the modification of the Heisenberg uncertainty relations, known as
\textit{Generalized Uncertainty Principle} (GUP)\cite{Xiang
Li,RenQinChun,GAC1,Mod}. The modified inequality takes the form%
\begin{equation}
\Delta x\Delta p\geq\hbar+\frac{\lambda_{p}^{2}}{\hbar}\left(  \Delta
p\right)  ^{2},
\end{equation}
where $\hbar$ is the Planck constant and $\lambda_{p}$ is the Planck length.
The interesting point regards exactly the modified number of quantum states,
which is changed into%
\begin{equation}
\frac{d^{3}xd^{3}p}{\left(  2\pi\hbar\right)  ^{3}\left(  1+\lambda
p^{2}\right)  ^{3}}.\label{eqn:states}%
\end{equation}
When $\lambda=0$, the formula reduces to the ordinary counting of quantum
states. If Eq.$\left(  \ref{eqn:states}\right)  $ is used for computing the
entropy, the brick wall can be removed\cite{Xiang Li,RenQinChun}. Note that
GUP and MDRs modifications are strictly connected\cite{CG}. This suggest that
we could use MDRs to remove the brick wall. MDRs have a deep impact also when
the background is curved. Indeed, the analysis of Magueijo and
Smolin\cite{MagSmo} shows that the energy-momentum tensor and the Einstein
equations are replaced by a one parameter family of equations%
\begin{equation}
G_{\mu\nu}\left(  E\right)  =8\pi G\left(  E\right)  T_{\mu\nu}\left(
E\right)  +g_{\mu\nu}\Lambda\left(  E\right)  ,
\end{equation}
where $G\left(  E\right)  $ is an energy dependent Newton's constant, defined
so that $G\left(  0\right)  $ is the physical Newton's constant. Similarly we
have an energy dependent cosmological constant $\Lambda\left(  E\right)  $. In
this context, the modified Friedmann-Robertson-Walker line element changes
into\cite{MagSmo}%
\begin{equation}
ds^{2}\left(  E\right)  =-\frac{dt^{2}}{g_{1}^{2}\left(  E\right)  }%
+\frac{a^{2}\left(  t\right)  }{g_{2}^{2}\left(  E\right)  }g_{ij}dx^{i}%
dx^{j},\label{line1}%
\end{equation}
where $g_{ij}$ represents the spatially homogeneous and isotropic metric of a
sphere (positive curvature $K=1$), pseudo-sphere (with negative curvature
$K=-1$), or euclidean space ( $K=0$, so that $g_{ij}=\delta_{ij}$). Note that
the metric coefficients are energy dependent. When a gravitational background
has a structure like the one in the line element $\left(  \ref{line1}\right)
$, we have a \textquotedblleft\textit{rainbow metric\textquotedblright.} When
the Schwarzschild line is examined, the related rainbow metric reads%
\begin{equation}
ds^{2}\left(  E\right)  =-\left(  1-\frac{2MG\left(  0\right)  }{r}\right)
\frac{d\tilde{t}^{2}}{g_{1}^{2}\left(  E\right)  }+\frac{d\tilde{r}^{2}%
}{\left(  1-\frac{2MG\left(  0\right)  }{r}\right)  g_{2}^{2}\left(  E\right)
}+\frac{\tilde{r}^{2}}{g_{2}^{2}\left(  E\right)  }\left(  d\theta^{2}%
+\sin^{2}\theta d\phi^{2}\right)  .\label{line2}%
\end{equation}
We expect the functions $g_{1}\left(  E\right)  $ and $g_{2}\left(  E\right)
$ modify the UV behavior in the same way as GUP and Noncommutative geometry
do, respectively. Since the form of $g_{1}\left(  E\right)  $ and
$g_{2}\left(  E\right)  $ is unknown and they have to obey the property
$\left(  \ref{prop}\right)  $, we have a large amount of arbitrariness in
fixing the dependence on $E/E_{P}$, even if some specific choices have been
proposed by G. Amelino-Camelia et al.\cite{Amelino et Al.,VSL MDR} in the
context of black hole thermodynamics. The rest of the paper is structured as
follows, in section \ref{p2} we compute the free energy and we examine its
contribution near the horizon, in section \ref{p3} we compute the relevant
thermodynamical quantities. We summarize and conclude in section \ref{p4}.
Units in which $\hbar=c=k=1$ are used throughout the paper.

\section{W.K.B. Approach with MDRs}

\label{p2}Instead of working with the background of form \ref{line1} or
\ref{line2}, we adopt the following general aspect for the line element,
useful for spherically symmetric problems\cite{Visser}%
\begin{equation}
ds^{2}=-\exp\left(  -2\Lambda\left(  r\right)  \right)  \left(  1-\frac
{b\left(  r\right)  }{r}\right)  \frac{dt^{2}}{g_{1}^{2}\left(  E/E_{P}%
\right)  }+\frac{dr^{2}}{\left(  1-\frac{b\left(  r\right)  }{r}\right)
g_{2}^{2}\left(  E/E_{P}\right)  }+\frac{r^{2}}{g_{2}^{2}\left(
E/E_{P}\right)  }d\Omega^{2}.\label{e31}%
\end{equation}
Usually, this kind of metric is adopted for the description of wormholes.
However, it is quite general to include as special cases the Schwarzschild,
Reissner-Nordstr{\"{o}}m and de Sitter and Anti-de Sitter geometries, or any
combination of these. The function $b\left(  r\right)  $ will be referred to
as the \textquotedblleft shape function\textquotedblright. The shape function
may be thought of as specifying the shape of the spatial slices. On the other
hand, $\Lambda\left(  r\right)  $ will be referred to as the \textquotedblleft
redshift function\textquotedblright\ that describes how far the total
gravitational redshift deviates from that implied by the shape function.
Without loss of generality we can fix the value of $\Lambda\left(  r\right)  $
at infinity such that $\Lambda\left(  \infty\right)  =0$. If the equation
$b\left(  r_{w}\right)  =r_{w}$ is satisfied for some values of $r$, then we
say that the points $r_{w}$ are horizons for the metric $\left(
\ref{e31}\right)  $. For the outermost horizon one has $\forall r>r_{w}$ that
$b\left(  r\right)  <r$. Consequently $b^{\prime}\left(  r_{w}\right)  \leq1$.
We will fix our attention to the $b^{\prime}\left(  r_{w}\right)  <1$ case
only. The anomalous case $b^{\prime}\left(  r_{w}\right)  =1$ can be thought
as describing extreme black holes where an inner and outer horizons are merged
and will not be considered here. For a spherically symmetric system the
surface gravity is computed via
\begin{equation}
\kappa_{w}=\lim_{r\rightarrow r_{w}}\left\{  \frac{1}{2}\frac{\partial
_{r}g_{tt}}{\sqrt{g_{tt}g_{rr}}}\right\}
\end{equation}
and for the metric $\left(  \ref{e31}\right)  $, we get
\begin{equation}
\kappa_{w}=\lim_{r\rightarrow r_{w}}\frac{1}{2}\left\{  \frac{\exp\left(
-\Lambda\left(  r\right)  \right)  }{h\left(  E/E_{P}\right)  }\left[
-2\Lambda^{\prime}\left(  r\right)  \left(  1-\frac{b\left(  r\right)  }%
{r}\right)  +\frac{b\left(  r\right)  }{r^{2}}-\frac{b^{\prime}\left(
r\right)  }{r}\right]  \right\}  ,\label{kw}%
\end{equation}
with%
\begin{equation}
h\left(  E/E_{P}\right)  =\frac{g_{1}\left(  E/E_{P}\right)  }{g_{2}\left(
E/E_{P}\right)  }.\label{h(E)}%
\end{equation}
By assuming that $\Lambda\left(  r_{w}\right)  $ and $\Lambda^{\prime}\left(
r_{w}\right)  $ are both finite we obtain that
\begin{equation}
\kappa_{w}=\frac{1}{2r_{w}h\left(  E\right)  }\exp\left(  -\Lambda\left(
r_{w}\right)  \right)  \left[  1-b^{\prime}\left(  r_{w}\right)  \right]
,\label{e32}%
\end{equation}
where, in the proximity of the throat we have approximated $1-b\left(
r\right)  /r$ with%
\begin{equation}
1-\frac{b\left(  r\right)  }{r}=\frac{r-r_{w}}{r_{w}}\text{ }\left[
1-b^{\prime}\left(  r_{w}\right)  \right]  .\label{e32a}%
\end{equation}
Now that the geometrical framework has been set up, we begin with a real
massless scalar field described by the action\footnote{See also
Ref.\cite{MukohyamaIsrael} for a derivation of the brick wall in the Boulware
state.}
\begin{equation}
I=-\frac{1}{2}\int d^{4}x\sqrt{-g}\left[  g^{\mu\nu}\partial_{\mu}\phi
\partial_{\nu}\phi\right]
\end{equation}
in the background geometry of Eq.$\left(  \ref{e31}\right)  $ whose
Euler-Lagrange equations are%
\begin{equation}
\frac{1}{\sqrt{-g}}\partial_{\mu}\left(  \sqrt{-g}g^{\mu\nu}\partial_{\nu
}\right)  \phi=0.
\end{equation}
If $\phi$ has the separable form
\begin{equation}
\phi\left(  t,r,\theta,\varphi\right)  =\exp\left(  -iEt\right)  Y_{lm}%
(\theta,\varphi)f\left(  r\right)  ,
\end{equation}
then the equation for $f\left(  r\right)  $ reads%
\begin{equation}
\left[  \frac{g_{2}^{2}\left(  E\right)  \exp\left(  \Lambda\left(  r\right)
\right)  }{r^{2}}\partial_{r}\left(  r^{2}\exp\left(  -\Lambda\left(
r\right)  \right)  \left(  1-\frac{b\left(  r\right)  }{r}\right)
\partial_{r}\right)  -\frac{l(l+1)}{r^{2}}+\frac{E_{nl}^{2}g_{1}^{2}\left(
E\right)  \exp\left(  2\Lambda\left(  r\right)  \right)  }{1-\frac{b\left(
r\right)  }{r}}\right]  f_{nl}=0,\label{e33}%
\end{equation}
where $Y_{lm}(\theta,\varphi)$ is the usual spherical harmonic function. In
order to make our system finite let us suppose that two mirror-like boundaries
are placed at $r=r_{1}$ and $r=R$ with $R\gg r_{1},$ $r_{1}>r_{w}$ and
consider Dirichlet boundary conditions $f_{nl}(r_{1})=f_{nl}(R)=0$. We also
assume the set of real functions $\{f_{nl}(r)\}$ ($n=1,2,\cdots$), defined by
Eq.$\left(  \ref{e33}\right)  $, be complete with respect to the space of
$L_{2}$-functions on the interval $r_{1}\leq r\leq R$ for each $l$. The
positive constant $\omega_{nl}$ is defined as the corresponding eigenvalue. In
order to use the WKB approximation, we define an r-dependent radial wave
number $k(r,l,E)$
\begin{equation}
k_{r}^{2}(r,l,E)\equiv\frac{1}{\left(  1-\frac{b\left(  r\right)  }{r}\right)
}\left[  \exp\left(  2\Lambda\left(  r\right)  \right)  \frac{E^{2}%
h^{2}\left(  E\right)  }{\left(  1-\frac{b\left(  r\right)  }{r}\right)
}-\frac{l(l+1)}{r^{2}}\right]  ,\label{squarekr}%
\end{equation}
where we have used Eq.$\left(  \ref{h(E)}\right)  $. The number of modes with
frequency less than $E$ is given approximately by
\begin{equation}
\tilde{g}(E)=\int_{0}^{l_{max}}\nu(l{,E})(2l+1)dl,
\end{equation}
where $\nu(l,E)$ is the number of nodes in the mode with $(l,E)$:
\begin{equation}
\nu(l,E)=\frac{1}{\pi}\int_{r_{w}}^{R}\sqrt{k^{2}(r,l,E)}dr.
\end{equation}
Here it is understood that the integration with respect to $r$ and $l$ is
taken over those values which satisfy $r_{w}\leq r\leq R$ and $k^{2}%
(r,l,E)\geq0$. Thus, from Eq.$\left(  \ref{squarekr}\right)  $ we get%
\[
\frac{d\tilde{g}(E)}{dE}=\int\frac{\partial\nu(l{,E})}{\partial E}%
(2l+1)dl=\frac{1}{\pi}\int_{r_{w}}^{R}dr\int_{0}^{l_{max}}dl(2l+1)\frac
{\exp\left(  2\Lambda\left(  r\right)  \right)  }{\sqrt{k^{2}(r,l,E)}}%
\frac{\left(  Eh^{2}\left(  E\right)  +E^{2}h\left(  E\right)  h^{\prime
}\left(  E\right)  \right)  }{\left(  1-\frac{b\left(  r\right)  }{r}\right)
^{2}}%
\]%
\begin{equation}
=\frac{2}{\pi}Eh\left(  E\right)  \left(  Eh^{2}\left(  E\right)
+E^{2}h\left(  E\right)  h^{\prime}\left(  E\right)  \right)  \int_{r_{w}}%
^{R}dr\frac{\exp\left(  3\Lambda\left(  r\right)  \right)  }{\left(
1-\frac{b\left(  r\right)  }{r}\right)  ^{2}}r^{2}=\frac{2}{\pi}\frac{d}%
{dE}\left(  \frac{1}{3}E^{3}h^{3}\left(  E\right)  \right)  \int_{r_{w}}%
^{R}dr\frac{\exp\left(  3\Lambda\left(  r\right)  \right)  }{\left(
1-\frac{b\left(  r\right)  }{r}\right)  ^{2}}r^{2}.\label{states}%
\end{equation}
The free energy is given approximately by
\begin{equation}
F=\frac{1}{\beta}\int_{0}^{\infty}\ln\left(  1-e^{-\beta E}\right)
\frac{d\tilde{g}(E)}{dE}dE,
\end{equation}
where $\beta$ is the inverse temperature measured at infinity. Then from
Eq.$\left(  \ref{states}\right)  $, we find
\begin{equation}
F=\frac{2}{\pi}\frac{1}{\beta}\int_{0}^{\infty}\ln\left(  1-e^{-\beta
E}\right)  dEE^{2}h^{2}\left(  E\right)  \left(  h\left(  E\right)
+Eh^{\prime}\left(  E\right)  \right)  \int_{r_{w}}^{R}drr^{2}\frac
{\exp\left(  3\Lambda\left(  r\right)  \right)  }{\left(  1-\frac{b\left(
r\right)  }{r}\right)  ^{2}}%
\end{equation}%
\begin{equation}
=\frac{2}{\pi}\frac{1}{\beta}\int_{0}^{\infty}\ln\left(  1-e^{-\beta
E}\right)  \frac{d}{dE}\left(  \frac{1}{3}E^{3}h^{3}\left(  E\right)  \right)
dE\int_{r_{w}}^{R}drr^{2}\frac{\exp\left(  3\Lambda\left(  r\right)  \right)
}{\left(  1-\frac{b\left(  r\right)  }{r}\right)  ^{2}}.
\end{equation}
It is convenient to divide the free energy into two pieces denoted by%
\begin{equation}
F_{r_{w}}=\frac{2}{\pi}\frac{1}{\beta}\int_{0}^{\infty}\ln\left(  1-e^{-\beta
E}\right)  dEE^{2}h^{2}\left(  E\right)  \left(  h\left(  E\right)
+Eh^{\prime}\left(  E\right)  \right)  \int_{r_{w}}^{r_{1}}drr^{2}\frac
{\exp\left(  3\Lambda\left(  r\right)  \right)  }{\left(  1-\frac{b\left(
r\right)  }{r}\right)  ^{2}}%
\end{equation}
and%
\begin{equation}
F_{R}=\frac{2}{\pi}\frac{1}{\beta}\int_{0}^{\infty}\ln\left(  1-e^{-\beta
E}\right)  dEE^{2}h^{2}\left(  E\right)  \left(  h\left(  E\right)
+Eh^{\prime}\left(  E\right)  \right)  \int_{r_{1}}^{R}drr^{2}\frac
{\exp\left(  3\Lambda\left(  r\right)  \right)  }{\left(  1-\frac{b\left(
r\right)  }{r}\right)  ^{2}}.
\end{equation}
Assuming that $\Lambda\left(  r\right)  <\infty,$ $\forall r\in\left[
r_{w},+\infty\right)  $, $F_{R}$ is dominated by large volume effects for
large $R$. This is particular evident in the case when $b\left(  r\right)
<r$. Indeed, we get for the radial dependent integral%
\begin{equation}
F_{R}\sim\frac{2}{\pi\beta}\frac{R^{3}}{3}\int_{0}^{\infty}\ln\left(
1-e^{-\beta E}\right)  \frac{d}{dE}\left(  \frac{1}{3}E^{3}h^{3}\left(
E\right)  \right)  dE.
\end{equation}
This term will give the contribution to the entropy of a homogeneous quantum
gas in flat space at a uniform temperature $T$ when a MDR of the form $\left(
\ref{mdisp}\right)  $ is considered. We will not examine this large volume
contribution here. To study $F_{r_{w}}$, we use Eq.$\left(  \ref{e32a}\right)
$, then the radial part of $F_{r_{w}}$ becomes divergent in proximity of
$r_{w}$. This ultraviolet divergence has been cured by 't Hooft, who
introduced a \textquotedblleft\textit{brick wall }$r_{0}$\textquotedblright%
\ proportional to $l_{P}^{2}$. This is obtained by keeping only the leading
divergence in $F_{r_{w}}$ and introducing the proper distance from the throat%
\begin{equation}
\alpha=\int_{r_{w}}^{r_{w}+r_{0}}\frac{dr}{\sqrt{1-\frac{b\left(  r\right)
}{r}}}=\frac{2\sqrt{r_{0}}}{\sqrt{\frac{1-b^{\prime}\left(  r_{w}\right)
}{r_{w}}}}.
\end{equation}
Then $F_{r_{w}}$ becomes%
\begin{equation}
\int_{r_{w}+r_{0}}^{r_{1}}drr^{2}\frac{\exp\left(  3\Lambda\left(  r\right)
\right)  }{\left(  1-\frac{b\left(  r\right)  }{r}\right)  ^{2}}\simeq
4r_{w}^{5}\frac{\exp\left(  3\Lambda\left(  r_{w}\right)  \right)  }{\left(
1-b^{\prime}\left(  r_{w}\right)  \right)  ^{3}}\frac{1}{\alpha^{2}}%
\simeq4r_{w}^{5}\frac{\exp\left(  3\Lambda\left(  r_{w}\right)  \right)
}{\left(  1-b^{\prime}\left(  r_{w}\right)  \right)  ^{3}}\frac{1}{l_{P}^{2}}.
\end{equation}
Nevertheless, since spacetime is modified by a \textquotedblleft%
\textit{rainbow metric}\textquotedblright, it is quite natural that even the
\textquotedblleft\textit{brick wall}\textquotedblright\ is affected by this
distortion. To see such an effect, we perform the radial integration in
$F_{r_{w}}$, to obtain%
\begin{equation}
\int_{r_{w}+r_{0}}^{r_{1}}drr^{2}\frac{\exp\left(  3\Lambda\left(  r\right)
\right)  }{\left(  1-\frac{b\left(  r\right)  }{r}\right)  ^{2}}=\int
_{r_{w}+r\left(  E/E_{P}\right)  }^{r_{1}}drr^{2}\frac{\exp\left(
3\Lambda\left(  r\right)  \right)  }{\left(  1-\frac{b\left(  r\right)  }%
{r}\right)  ^{2}}\simeq r_{w}^{4}\frac{\exp\left(  3\Lambda\left(
r_{w}\right)  \right)  }{\left(  1-b^{\prime}\left(  r_{w}\right)  \right)
^{2}}\frac{1}{r\left(  E/E_{P}\right)  },\label{rw}%
\end{equation}
where we have assumed that, in proximity of the throat the brick wall is no
longer a constant but it becomes a function of $E/E_{P}$\footnote{In
Ref.\cite{RSS}, it has been introduced a radial dependent cut-off by setting%
\begin{equation}
\sqrt{1-\frac{r_{w}}{r_{E}}\left(  1-b^{\prime}\left(  r_{w}\right)  \right)
\Lambda}=E.
\end{equation}
This leads to%
\begin{equation}
r_{E}=r_{w}\frac{\left(  1-b^{\prime}\left(  r_{w}\right)  \right)
\Lambda^{2}}{\Lambda^{2}\left(  1-b^{\prime}\left(  r_{w}\right)  \right)
-E^{2}}\quad{\rightarrow\quad}r_{w},
\end{equation}
when $\Lambda\rightarrow\infty$. Thus, the r\^{o}le of $\Lambda$ is analogous
to the brick wall.}. Plugging Eq.$\left(  \ref{rw}\right)  $ into $F_{r_{w}}$,
we obtain%
\begin{equation}
F_{r_{w}}=\frac{2r_{w}^{4}}{\pi\beta}\frac{\exp\left(  3\Lambda\left(
r_{w}\right)  \right)  }{\left(  1-b^{\prime}\left(  r_{w}\right)  \right)
^{2}}\int_{0}^{\infty}\frac{\ln\left(  1-\exp\left(  -\beta E\right)  \right)
}{r\left(  E/E_{P}\right)  }\frac{d}{dE}\left(  \frac{1}{3}E^{3}h^{3}\left(
E/E_{P}\right)  \right)  dE.\label{Frw_00}%
\end{equation}
After an integration by parts, one gets%
\begin{equation}
F_{r_{w}}=-\frac{C_{r_{w}}}{3\beta}\int_{0}^{\infty}E^{3}h^{3}\left(
E/E_{P}\right)  \frac{d}{dE}\left[  \frac{\ln\left(  1-\exp\left(  -\beta
E\right)  \right)  }{r\left(  E/E_{P}\right)  }\right]  dE,\label{Frw}%
\end{equation}
where $h\left(  E/E_{P}\right)  $ is chosen in such a way to allow the
convergence when $E/E_{P}\rightarrow\infty$ and%
\begin{equation}
C_{r_{w}}=\frac{2r_{w}^{4}}{\pi}\frac{\exp\left(  3\Lambda\left(
r_{w}\right)  \right)  }{\left(  1-b^{\prime}\left(  r_{w}\right)  \right)
^{2}}.\label{e33a}%
\end{equation}
Since one is interested to an elimination of the \textquotedblleft%
\textit{brick wall}\textquotedblright\ when $E/E_{P}\rightarrow0$ we assume,
without loss of generality, that%
\begin{equation}
r\left(  E/E_{P}\right)  =r_{w}\sigma\left(  E/E_{P}\right)  ,\label{r(E)}%
\end{equation}
with%
\begin{equation}
\sigma\left(  E/E_{P}\right)  \rightarrow0,\qquad E/E_{P}\rightarrow
0.\label{r(E)sigma}%
\end{equation}
Plugging Eq.$\left(  \ref{r(E)}\right)  $ into Eq.$\left(  \ref{Frw}\right)
$, we obtain%
\begin{equation}
F_{r_{w}}=-\frac{C_{r_{w}}}{3\beta r_{w}}\int_{0}^{\infty}E^{3}h^{3}\left(
E\right)  \frac{d}{dE}\left[  \frac{\ln\left(  1-\exp\left(  -\beta E\right)
\right)  }{\sigma\left(  E/E_{P}\right)  }\right]  dE
\end{equation}%
\begin{equation}
=-\frac{C_{r_{w}}}{3\beta r_{w}}\int_{0}^{\infty}\frac{E^{3}h^{3}\left(
E\right)  }{\sigma\left(  E/E_{P}\right)  }\left[  \frac{\beta}{\left(
\exp\left(  \beta E\right)  -1\right)  }-\frac{\ln\left(  1-\exp\left(  -\beta
E\right)  \right)  }{E_{P}\sigma\left(  E/E_{P}\right)  }\sigma^{\prime
}\left(  E/E_{P}\right)  \right]  dE,\label{Frw_0}%
\end{equation}
where the prime means that we are computing the derivative with respect to the
argument. We can see what happens to the free energy $F_{r_{w}}$ for some
specific forms of $g_{1}\left(  E/E_{P}\right)  $ and $g_{2}\left(
E/E_{P}\right)  $. One popular choice is given by%
\begin{equation}
g_{1}\left(  E/E_{P}\right)  =1-\eta\left(  E/E_{P}\right)  ^{n}%
\qquad\text{and}\qquad g_{2}\left(  E/E_{P}\right)  =1,
\end{equation}
where $\eta$ is a dimensionless parameter and $n$ is an integer\cite{g1g2}.
Thus the form of $h\left(  E/E_{P}\right)  $ is%
\begin{equation}
h\left(  E/E_{P}\right)  =1-\eta\left(  E/E_{P}\right)  ^{n}.
\end{equation}
Nevertheless, the above choice does not allow the integration by parts in
Eq.$\left(  \ref{Frw_00}\right)  $ and therefore will be discarded. Thus the
choice of the possible forms of $g_{1}\left(  E/E_{P}\right)  $ and
$g_{2}\left(  E/E_{P}\right)  $ is strongly restricted by convergence criteria
and by the property $\left(  \ref{r(E)sigma}\right)  $. One good candidate for
the convergence is%
\begin{equation}
h\left(  E/E_{P}\right)  =\exp\left(  -\frac{E}{E_{P}}\right)  .\label{hexp}%
\end{equation}
In section \ref{p3}, we will discuss some choices for $\sigma\left(
E/E_{P}\right)  $.

\section{Entropy and Total Energy with MDRs}

\label{p3}The property described in $\left(  \ref{r(E)sigma}\right)  $ shows
that the divergence on the throat is represented by $\sigma\left(
E/E_{P}\right)  $. However, the distortion due to the rainbow metric in
Eq.$\left(  \ref{Frw_0}\right)  $ introduces a term $E^{3}$ which partially
cures the throat divergence. As an illustrative example, we consider the
following form for
\begin{equation}
\sigma\left(  E/E_{P}\right)  =h^{\delta}\left(  E/E_{P}\right)  \left(
\frac{E}{E_{P}}\right)  ^{\alpha} \label{sigma(E)}%
\end{equation}
which, of course do not exhaust the possible candidates for MDRs. We have two
interesting cases:

\begin{description}
\item[a)] $\delta=0;$ $\alpha>0$

\item and

\item[b)] $\delta>0;$ $\alpha>0$.
\end{description}

With choice \textbf{a)} Eq.$\left(  \ref{Frw}\right)  $ becomes%
\begin{equation}
F_{r_{w}}=-\frac{C_{r_{w}}E_{P}^{\alpha}}{3\beta r_{w}}\int_{0}^{\infty}%
E^{3}\exp\left(  -3E/E_{P}\right)  \left[  \frac{\beta}{\left(  \exp\left(
\beta E\right)  -1\right)  E^{\alpha}}-\frac{\alpha\ln\left(  1-\exp\left(
-\beta E\right)  \right)  }{E^{\alpha+1}}\right]  dE. \label{Frwa}%
\end{equation}
The second integral considerably simplifies for $\alpha=2.$ From the
Appendix\ref{Appe}, we find%
\begin{equation}
F_{r_{w}}=-\frac{2C_{r_{w}}E_{P}^{2}}{3\beta^{2}r_{w}}\left[  \zeta\left(
2,1+\frac{3}{\beta E_{P}}\right)  +\frac{\beta E_{P}}{3}\left(  \gamma
+\Psi\left(  1+\frac{3}{\beta E_{P}}\right)  \right)  \right]  ,
\end{equation}
where $\zeta\left(  s,\nu\right)  $ is the Hurwitz zeta function,
$\Gamma\left(  x\right)  $ is the gamma function and $\Psi\left(  x\right)  $
is the digamma function. Since the Hurwitz zeta function obeys the relation%
\begin{equation}
\frac{d}{d\nu}\zeta\left(  s,\nu\right)  =-s\zeta\left(  s+1,\nu\right)  ,
\end{equation}
we can easily compute the other relevant thermodynamic quantities. The total
energy $U$ is defined by%
\[
U=\frac{\partial\left(  \beta F_{r_{w}}\right)  }{\partial\beta}%
=-\frac{2C_{r_{w}}E_{P}^{2}}{3r_{w}}\frac{\partial}{\partial\beta}\left[
\frac{1}{\beta}\zeta\left(  2,1+\frac{3}{\beta E_{P}}\right)  +\frac{E_{P}}%
{3}\left(  \gamma+\Psi\left(  1+\frac{3}{\beta E_{P}}\right)  \right)
\right]
\]%
\begin{equation}
=\frac{2C_{r_{w}}E_{P}^{2}}{3r_{w}\beta^{2}}\left[  \zeta\left(  2,1+\frac
{3}{\beta E_{P}}\right)  -\zeta\left(  3,1+\frac{3}{\beta E_{P}}\right)
\left(  \frac{6}{\beta E_{P}}\right)  +\Psi^{\prime}\left(  1+\frac{3}{\beta
E_{P}}\right)  \right]
\end{equation}
and the entropy $S$ is%
\[
S=\beta^{2}\frac{\partial F_{r_{w}}}{\partial\beta}=-\beta^{2}\frac{2C_{r_{w}%
}E_{P}^{2}}{3r_{w}}\frac{\partial}{\partial\beta}\left[  \frac{1}{\beta^{2}%
}\zeta\left(  2,1+\frac{3}{\beta E_{P}}\right)  +\frac{E_{P}}{3\beta}\left(
\gamma+\Psi\left(  1+\frac{3}{\beta E_{P}}\right)  \right)  \right]
\]%
\[
=\frac{2C_{r_{w}}E_{P}^{2}}{3r_{w}}\,\left[  \frac{2}{\beta}\zeta\left(
2,1+\frac{3}{\beta E_{P}}\right)  +\frac{E_{P}}{3}\left(  \gamma+\Psi\left(
1+\frac{3}{\beta E_{P}}\right)  \right)  \right.
\]%
\begin{equation}
\left.  -\zeta\left(  3,1+\frac{3}{\beta E_{P}}\right)  \left(  \frac{6}%
{\beta^{2}E_{P}}\right)  +\frac{1}{\beta}\Psi^{\prime}\left(  1+\frac{3}{\beta
E_{P}}\right)  \right]  .
\end{equation}
In the limit where $\beta E_{P}\gg1$, we can take the leading order of the
total energy $U$%
\begin{equation}
U=\frac{2C_{r_{w}}E_{P}^{2}}{3\beta^{2}r_{w}}\left[  \zeta\left(  2\right)
+\frac{\pi^{2}}{6}\right]  =\frac{2C_{r_{w}}E_{P}^{2}}{3\beta^{2}r_{w}}%
\frac{\pi^{2}}{3}=r_{w}^{3}\frac{\exp\left(  3\Lambda\left(  r_{w}\right)
\right)  }{\left(  1-b^{\prime}\left(  r_{w}\right)  \right)  ^{2}}%
\frac{4E_{P}^{2}}{9\beta^{2}}\pi
\end{equation}
and the entropy $S$%
\begin{equation}
S=\frac{2C_{r_{w}}E_{P}^{2}}{3r_{w}}\,\left[  \frac{2}{\beta}\zeta\left(
2\right)  +\frac{E_{P}}{3}\left(  \gamma-\gamma+\frac{\pi^{2}}{2\beta E_{P}%
}\right)  +\frac{\pi^{2}}{6\beta}\right]  =r_{w}^{3}\frac{\exp\left(
3\Lambda\left(  r_{w}\right)  \right)  }{\left(  1-b^{\prime}\left(
r_{w}\right)  \right)  ^{2}}\frac{8E_{P}^{2}}{9\beta}\pi,
\end{equation}
where we have used Eq.$\left(  \ref{e33a}\right)  $. Moreover, recalling the
expression for the surface gravity in the low energy limit, we get%
\begin{equation}
\kappa_{w}=\frac{1}{2r_{w}}\exp\left(  -\Lambda\left(  r_{w}\right)  \right)
\left[  1-b^{\prime}\left(  r_{w}\right)  \right]  . \label{kwl}%
\end{equation}
then $U$%
\begin{equation}
U=r_{w}^{2}\frac{\exp\left(  2\Lambda\left(  r_{w}\right)  \right)
}{1-b^{\prime}\left(  r_{w}\right)  }\frac{2E_{P}^{2}}{9\beta^{2}\kappa_{w}%
}\pi
\end{equation}
and the entropy $S$%
\begin{equation}
S=r_{w}^{2}\frac{\exp\left(  2\Lambda\left(  r_{w}\right)  \right)
}{1-b^{\prime}\left(  r_{w}\right)  }\frac{4E_{p}^{2}}{9\beta\kappa_{w}}\pi
\end{equation}
lead to%
\begin{equation}
U=r_{w}^{2}\frac{\exp\left(  2\Lambda\left(  r_{w}\right)  \right)
}{1-b^{\prime}\left(  r_{w}\right)  }\frac{E_{P}^{2}}{9\beta}%
\end{equation}
and the entropy $S$%
\begin{equation}
S=\frac{A_{r_{w}}E_{P}^{2}}{4}\frac{\exp\left(  2\Lambda\left(  r_{w}\right)
\right)  }{1-b^{\prime}\left(  r_{w}\right)  }\frac{2}{9\pi},
\end{equation}
where we have used%
\begin{equation}
\frac{1}{\beta}=T=\frac{\kappa_{w}}{2\pi}. \label{e35}%
\end{equation}
To recover the area law, we have to set%
\begin{equation}
\frac{\exp\left(  2\Lambda\left(  r_{w}\right)  \right)  }{1-b^{\prime}\left(
r_{w}\right)  }=\frac{9\pi}{2}.
\end{equation}
This corresponds to a changing of the time variable with respect to the
Schwarzschild time. The total energy becomes%
\begin{equation}
U=r_{w}^{2}\frac{\pi E_{P}^{2}}{2\beta}, \label{e36}%
\end{equation}
which in terms of the Schwarzschild radius $r_{w}=2MG$ and inverse Hawking
temperature $\beta=8\pi MG$ becomes%
\begin{equation}
U=4M^{2}G^{2}\frac{E_{P}^{2}}{16MG}=\frac{M}{4}.
\end{equation}
Note the discrepancy of a factor of $3/2$ with the 't Hooft result.

With the choice \textbf{b)}, $F_{r_{w}}$becomes%
\[
F_{r_{w}}=-\frac{C_{r_{w}}}{3\beta r_{w}}\int_{0}^{\infty}dEE^{3}h^{3}\left(
E\right)  \left[  \frac{E_{P}^{\alpha}\beta}{\left(  \exp\left(  \beta
E\right)  -1\right)  h^{\delta}\left(  E\right)  E^{\alpha}}\right.
\]%
\begin{equation}
\left.  -\frac{E_{P}^{\alpha}\ln\left(  1-\exp\left(  -\beta E\right)
\right)  }{h^{\delta}\left(  E/E_{P}\right)  E^{\alpha}E_{P}}\left(  \delta
E_{P}\frac{h^{\prime}\left(  E/E_{P}\right)  }{h\left(  E/E_{P}\right)
}+\alpha\frac{E_{P}}{E}\right)  \right]  .
\end{equation}
Since $h\left(  E\right)  $ assumes the form \ref{hexp}, we can further
simplify the above integral%
\begin{equation}
F_{r_{w}}=-\frac{C_{r_{w}}E_{P}^{\alpha}}{3\beta r_{w}}\int_{0}^{\infty
}dEE^{3-\alpha}h^{3-\delta}\left(  E\right)  \left[  \frac{\beta}{\left(
\exp\left(  \beta E\right)  -1\right)  }-\ln\left(  1-\exp\left(  -\beta
E\right)  \right)  \left(  -\frac{\delta}{E_{P}}+\frac{\alpha}{E}\right)
\right]  . \label{Frwb}%
\end{equation}
Fixing $\alpha=2$ and letting $\delta$ unspecified, we get%
\begin{equation}
F_{r_{w}}=-\frac{C_{r_{w}}E_{P}^{2}}{3\beta r_{w}}\left\{  \frac{2}{\beta
}\zeta\left(  2,1+\frac{3-\delta}{\beta E_{P}}\right)  +\frac{\delta E_{P}%
}{\left(  3-\delta\right)  ^{2}}\left[  \gamma+\Psi\left(  1+\frac{3-\delta
}{\beta E_{P}}\right)  \right]  +\delta\frac{\Psi^{\prime}\left(
1+\frac{3-\delta}{\beta E_{P}}\right)  }{\beta\left(  3-\delta\right)  }%
+\frac{\pi^{2}}{3\beta}\right\}  .
\end{equation}
where we have integrated by parts the last term of Eq.$\left(  \ref{Frwb}%
\right)  $. The value of $\delta=3$ has to be treated as a separate case,
which reduces $F_{r_{w}}$ to%
\begin{equation}
F_{r_{w}}=-\frac{C_{r_{w}}E_{P}^{2}}{3\beta r_{w}}\int_{0}^{\infty}dEE\left[
\frac{\beta}{\left(  \exp\left(  \beta E\right)  -1\right)  }+3\frac
{\ln\left(  1-\exp\left(  -\beta E\right)  \right)  }{E_{P}}-2\frac{\ln\left(
1-\exp\left(  -\beta E\right)  \right)  }{E}\right]  .
\end{equation}
After an integration by parts on the second and third term, we obtain%
\begin{equation}
F_{r_{w}}=-\frac{C_{r_{w}}E_{P}^{2}}{3\beta^{2}r_{w}}\int_{0}^{\infty
}dx\left[  \frac{3x^{2}}{2\beta E_{P}\left(  e^{x}-1\right)  }+\frac
{3x}{\left(  e^{x}-1\right)  }\right]  =-\frac{C_{r_{w}}E_{P}^{2}}{\beta
^{2}r_{w}}\left(  \frac{\pi^{2}}{6}+\frac{\zeta\left(  3\right)  }{\beta
E_{P}}\right)  .
\end{equation}
With the help of Eq.$\left(  \ref{kwl}\right)  $ and in the approximation
$\beta E_{P}\gg1$, the relevant thermodynamical quantities are
\begin{equation}
U=\frac{C_{r_{w}}E_{P}^{2}}{6\beta^{2}r_{w}}\pi^{2}=r_{w}^{2}\frac{\exp\left(
2\Lambda\left(  r_{w}\right)  \right)  }{1-b^{\prime}\left(  r_{w}\right)
}\frac{E_{P}^{2}}{12\beta}%
\end{equation}
and%
\begin{equation}
S=\frac{\pi^{2}C_{r_{w}}E_{P}^{2}}{3\beta r_{w}}=\frac{A_{r_{w}}}{4}E_{P}%
^{2}\frac{\exp\left(  2\Lambda\left(  r_{w}\right)  \right)  }{1-b^{\prime
}\left(  r_{w}\right)  }\frac{1}{6\pi}.
\end{equation}
Even in this case, to recover the area law, we have to set%
\begin{equation}
\frac{\exp\left(  2\Lambda\left(  r_{w}\right)  \right)  }{1-b^{\prime}\left(
r_{w}\right)  }=6\pi
\end{equation}
and the total energy becomes%
\begin{equation}
U=\frac{C_{r_{w}}E_{P}^{2}}{6\beta^{2}r_{w}}\pi^{2}=r_{w}^{2}\frac{\pi
E_{P}^{2}}{2\beta}.=\frac{M}{4}.
\end{equation}
In terms of the inverse Hawking temperature and of the Schwarzschild radius,
we get%
\begin{equation}
U=\frac{M}{4}%
\end{equation}
differing from the 't Hooft result by a factor of $3/2$. When $\delta\neq3$,
from Eq.$\left(  \ref{Udelta}\right)  $ of Appendix \ref{Appen}, we extract
the form of the internal energy%
\begin{equation}
U=r_{w}^{2}\frac{\exp\left(  2\Lambda\left(  r_{w}\right)  \right)
}{1-b^{\prime}\left(  r_{w}\right)  }\frac{E_{P}^{2}}{6\beta}\left[
\frac{\delta}{3-\delta}+\frac{2}{3}\right]  ,
\end{equation}
while for the entropy from Eq.$\left(  \ref{Sdelta}\right)  $, we can write%
\begin{equation}
S=r_{w}^{2}\frac{\exp\left(  2\Lambda\left(  r_{w}\right)  \right)
}{1-b^{\prime}\left(  r_{w}\right)  }\frac{E_{P}^{2}}{3}\left[  \frac{\delta
}{6\left(  3-\delta\right)  }+\frac{2}{3}\right]  =\frac{A_{r_{w}}}{4}%
E_{P}^{2}\frac{\exp\left(  2\Lambda\left(  r_{w}\right)  \right)
}{1-b^{\prime}\left(  r_{w}\right)  }\frac{1}{3\pi}\left[  \frac{\delta
}{6\left(  3-\delta\right)  }+\frac{2}{3}\right]  .
\end{equation}
If we set%
\begin{equation}
\frac{\exp\left(  2\Lambda\left(  r_{w}\right)  \right)  }{1-b^{\prime}\left(
r_{w}\right)  }=9\pi\left[  \frac{\delta}{2\left(  3-\delta\right)
}+2\right]  ^{-1},
\end{equation}
we recover the area law and the internal energy becomes%
\begin{equation}
U=r_{w}^{2}\pi\frac{E_{P}^{2}}{\beta}\left[  \frac{\delta+6}{12-\delta
}\right]  .
\end{equation}

\section{Summary and discussion}

\label{p4}In this paper, we have examined the possibility that MDRs alter
space time so deeply that even the black hole physics close to the horizon is
affected. In particular, we have examined the possibility that MDRs cure the
high frequency pathology known as \textit{brick wall}. We have based our
approach using a particular form of MDRs known as \textquotedblleft%
\textit{gravity's rainbow}\textquotedblright\textit{ }which introduces two
unknown functions $g_{1}\left(  E/E_{P}\right)  $ and $g_{2}\left(
E/E_{P}\right)  $ in the background metric. These two unknown functions have
the property $\left(  \ref{prop}\right)  $ which means that for low energy, we
are dealing with ordinary gravity. Since we are interested in the entropy and
energy computation respectively, we have used the WKB approximation which
introduces a density of states dependent on $g_{1}\left(  E/E_{P}\right)  $
and $g_{2}\left(  E/E_{P}\right)  $. Although the form of the
\textquotedblleft\textit{rainbow}\textquotedblright\ functions $g_{1}\left(
E/E_{P}\right)  $ and $g_{2}\left(  E/E_{P}\right)  $ is undetermined, we can
receive hints from the function $h\left(  E/E_{P}\right)  $ in Eq.$\left(
\ref{h(E)}\right)  $, representing the ratio between $g_{1}\left(
E/E_{P}\right)  $ and $g_{2}\left(  E/E_{P}\right)  $. Indeed, it appears that
in order to have finite values of the free energy, $h\left(  E/E_{P}\right)  $
must be of a form to allow the convergence of the integral in Eq.$\left(
\ref{Frw}\right)  $. Note that the \textquotedblleft\textit{rainbow
metric}\textquotedblright\ $\left(  \ref{prop}\right)  $ affects the free
energy UV\ by means of the term $\frac{1}{3}E^{3}h^{3}\left(  E/E_{P}\right)
$ in the density of states and plays a central role in the elimination of the
\textit{brick wall}. A key point comes from the assumption that the
\textit{brick wall} can be energy dependent. Unfortunately, an explicit
expression of such a dependence cannot be extracted at this level. It is
likely that such an expression be revealed by a further examination of
Einstein's field equations in the context of a \textquotedblleft%
\textit{gravity's rainbow}\textquotedblright. A tentative of computing a form
of $r\left(  E/E_{P}\right)  $ has been suggested in Ref.\cite{RSS}, even if a
sharp UV cutoff has been used . If one is tempted to perform a series
expansion at low energies%
\begin{equation}
r\left(  E/E_{P}\right)  =r^{\prime}\left(  E/E_{P}\right)  \frac{E}{E_{P}%
}+\ldots
\end{equation}
with the condition%
\begin{equation}
r\left(  E/E_{P}\right)  \rightarrow0\text{\qquad}\mathrm{when}\qquad
E/E_{P}\rightarrow0,
\end{equation}
one discovers form Eq.$\left(  \ref{Frw_0}\right)  $ that the result is
finite. However, this works only for a low energy limit. Therefore, it appears
to be advantageous to guess what families of functions can be used to have a
finite free energy and therefore the related entropy. Indeed, the choice
$\left(  \ref{sigma(E)}\right)  $ represents a good combination of functions
having the right properties for the elimination of the \textit{brick wall}. We
wish to remark that the disappearance of the \textit{brick wall} is due to the
presence of MDRs and not to the fact that $r\equiv r\left(  E/E_{P}\right)  $
close to the horizon. Of course one can choose a dependence on $E$ such that
the divergence disappears without invoking MDRs. Nevertheless, MDRs are a
consequence of a space time modification, which affect even the horizon
behavior. In a sense the simple modification of $r\rightarrow r\left(
E/E_{P}\right)  $ represents the quantum fluctuations of the throat. That a
modification of space time be a cure of divergences in entropy computation has
been shown with a specific model of space time foam\cite{RIJMPD} context,
where the degree of the divergence has been lowered to logarithmic\cite{RMPLA}%
. This result suggests that space time foam be in strong connection with MDRs.
How to explicitly realize such a connection will be the subject of a further
investigation\cite{Remo}.

\appendix{}

\section{Integrals}

\label{Appe}In this Appendix, we explicitly compute the integrals appearing in
Eq.$\left(  \ref{Frwa}\right)  $. We begin with%
\[
\int_{0}^{\infty}dE\frac{h^{\delta}\left(  E\right)  E^{3-\alpha}}{e^{\beta
E}-1}=\int_{0}^{\infty}dE\frac{\exp\left(  -\delta E/E_{P}\right)
E^{3-\alpha}}{e^{\beta E}-1}.
\]
By setting $\beta E=x$, we get%
\begin{equation}
\int_{0}^{\infty}dx\frac{\exp\left[  -\left(  1+\delta/\left(  \beta
E_{P}\right)  \right)  x\right]  x^{3-\alpha}}{1-e^{-x}}=\frac{\Gamma\left(
4-\alpha\right)  }{\beta^{4-\alpha}}\zeta\left(  4-\alpha,1+\frac{\delta
}{\beta E_{P}}\right)  , \label{a1}%
\end{equation}
where we have used the formula%
\[
\int_{0}^{\infty}dx\frac{x^{\nu-1}\exp\left(  -\mu x\right)  }{1-e^{-\beta x}%
}=\frac{\Gamma\left(  \nu\right)  }{\beta^{\nu}}\zeta\left(  \nu,\frac{\mu
}{\beta}\right)  \qquad%
\begin{array}
[c]{c}%
Re\mu>0\\
Re\nu>1
\end{array}
.
\]

The term%
\begin{equation}
\int_{0}^{\infty}dEEh^{3-\delta}\left(  E\right)  \ln\left(  1-e^{-\beta
E}\right)  =\int_{0}^{\infty}dEE\exp\left(  -\left(  3-\delta\right)
E/E_{P}\right)  \ln\left(  1-e^{-\beta E}\right)  , \label{h(E)ln}%
\end{equation}
can be cast into the form $\left(  \beta E=x\right)  $%
\[
=\frac{1}{\beta^{2}}\int_{0}^{\infty}dxx\exp\left(  -\frac{\left(
3-\delta\right)  x}{\beta E_{P}}\right)  \ln\left(  1-e^{-x}\right)
=-\frac{d}{d\mu}\left[  \frac{1}{\beta^{2}}\int_{0}^{\infty}dx\exp\left(
-\left(  \mu+\frac{\left(  3-\delta\right)  }{\beta E_{P}}\right)  x\right)
\ln\left(  1-e^{-x}\right)  \right]  _{|\mu=0}%
\]%
\begin{equation}
=-\frac{d}{d\mu}\left[  \frac{1}{\beta^{2}}\int_{0}^{\infty}dx\exp\left(
-\left(  a+\mu\right)  x\right)  \ln\left(  1-e^{-x}\right)  \right]
_{|\mu=0}=-\frac{d}{d\mu}\left[  \frac{1}{\beta^{2}}\int_{0}^{1}dt\left(
1-t\right)  ^{a+\mu-1}\ln\left(  t\right)  \right]  _{|\mu=0}%
\end{equation}
with%
\begin{equation}
a=\frac{3-\delta}{\beta E_{P}}.
\end{equation}
The result of the integration gives%
\begin{equation}
\int_{0}^{1}dt\left(  1-t\right)  ^{a+\mu-1}\ln\left(  t\right)  =B\left(
1,a+\mu\right)  \left[  \Psi\left(  1\right)  -\Psi\left(  1+a+\mu\right)
\right]
\end{equation}%
\begin{equation}
=\frac{\Gamma\left(  a+\mu\right)  }{\Gamma\left(  1+a+\mu\right)  }\left[
\Psi\left(  1\right)  -\Psi\left(  1+a+\mu\right)  \right]  =\frac{1}{a+\mu
}\left[  -\gamma-\Psi\left(  1+a+\mu\right)  \right]  ,
\end{equation}
where $B\left(  x,y\right)  $ is the Beta function, $\Psi\left(  x\right)  $
is the digamma function and where we have used the property%
\begin{equation}
\Gamma\left(  1+x\right)  =x\Gamma\left(  x\right)  .
\end{equation}
The integral \ref{h(E)ln} becomes%
\begin{equation}
-\frac{d}{d\mu}\left[  \frac{1}{\beta^{2}}\left(  \frac{1}{a+\mu}\left[
-\gamma-\Psi\left(  1+a+\mu\right)  \right]  \right)  \right]  _{|\mu=0}%
=\frac{1}{\beta^{2}}\left[  \left(  \frac{1}{a^{2}}\left[  \gamma+\Psi\left(
1+a\right)  \right]  \right)  +\frac{\Psi^{\prime}\left(  1+a\right)  }%
{a}\right]  ,
\end{equation}
where we have used the relation%
\begin{equation}
\int_{0}^{1}dxx^{\mu-1}\left(  1-x^{r}\right)  ^{\nu-1}\ln\left(  x\right)
=\frac{1}{r^{2}}B\left(  \frac{\mu}{r},\nu\right)  \left[  \Psi\left(
\frac{\mu}{r}\right)  -\Psi\left(  \frac{\mu}{r}+\nu\right)  \right]  \qquad%
\begin{array}
[c]{c}%
Re\mu>0\\
Re\nu>0\\
r>0
\end{array}
.
\end{equation}

\section{Case $\delta\neq3$}

\label{Appen}In this Appendix, we compute the internal energy $U$ and the
entropy $S$ for $\delta\neq3$. For the internal energy, we obtain%
\[
U=\frac{\partial\left(  \beta F_{r_{w}}\right)  }{\partial\beta}%
=\frac{C_{r_{w}}E_{P}^{2}}{3\beta^{2}r_{w}}\left[  2\zeta\left(
2,1+\frac{3-\delta}{\beta E_{P}}\right)  +\delta\frac{\Psi^{\prime}\left(
1+\frac{3-\delta}{\beta E_{P}}\right)  }{3-\delta}+\frac{\pi^{2}}{3}\right]
\]%
\begin{equation}
+\frac{C_{r_{w}}E_{P}^{2}}{3\beta r_{w}}\left(  \frac{3-\delta}{\beta^{2}%
E_{P}}\right)  \left\{  -4\zeta\left(  2,1+\frac{3-\delta}{\beta E_{P}%
}\right)  +\frac{\delta E_{P}}{\left(  3-\delta\right)  ^{2}}\Psi^{\prime
}\left(  1+\frac{3-\delta}{\beta E_{P}}\right)  +\delta\frac{\Psi
^{\prime\prime}\left(  1+\frac{3-\delta}{\beta E_{P}}\right)  }{\left(
3-\delta\right)  }\right\}  .
\end{equation}
In the limit $\beta E_{P}\gg1$, we find%
\begin{equation}
U\simeq\frac{C_{r_{w}}E_{P}^{2}}{3\beta^{2}r_{w}}\left[  2\zeta\left(
2\right)  +\delta\frac{\Psi^{\prime}\left(  1\right)  }{3-\delta}+\frac
{\pi^{2}}{3}\right]  =\frac{C_{r_{w}}E_{P}^{2}}{3\beta^{2}r_{w}}\pi^{2}\left[
\frac{\delta}{3-\delta}+\frac{2}{3}\right]  . \label{Udelta}%
\end{equation}
For the entropy, from the definition one gets%
\begin{equation}
F_{r_{w}}=-\frac{C_{r_{w}}E_{P}^{2}}{3\beta r_{w}}\left\{  \frac{2}{\beta
}\zeta\left(  2,1+\frac{3-\delta}{\beta E_{P}}\right)  +\frac{\delta E_{P}%
}{\left(  3-\delta\right)  ^{2}}\left[  \gamma+\Psi\left(  1+\frac{3-\delta
}{\beta E_{P}}\right)  \right]  +\delta\frac{\Psi^{\prime}\left(
1+\frac{3-\delta}{\beta E_{P}}\right)  }{\beta\left(  3-\delta\right)  }%
+\frac{\pi^{2}}{3\beta}\right\}  .
\end{equation}%
\[
S=\beta^{2}\frac{\partial F_{r_{w}}}{\partial\beta}=\frac{C_{r_{w}}E_{P}^{2}%
}{3r_{w}}\left[  \frac{4}{\beta}\zeta\left(  2,1+\frac{3-\delta}{\beta E_{P}%
}\right)  +2\delta\frac{\Psi^{\prime}\left(  1+\frac{3-\delta}{\beta E_{P}%
}\right)  }{\beta\left(  3-\delta\right)  }+\frac{2\pi^{2}}{3\beta}%
+\frac{\delta E_{P}}{\left(  3-\delta\right)  ^{2}}\left[  \gamma+\Psi\left(
1+\frac{3-\delta}{\beta E_{P}}\right)  \right]  \right]
\]%
\begin{equation}
+\frac{C_{r_{w}}E_{P}}{3\beta r_{w}}\left(  3-\delta\right)  \left\{
-\frac{4}{\beta}\zeta\left(  3,1+\frac{3-\delta}{\beta E_{P}}\right)
+\frac{\delta E_{P}}{\left(  3-\delta\right)  ^{2}}\Psi^{\prime}\left(
1+\frac{3-\delta}{\beta E_{P}}\right)  +\delta\frac{\Psi^{\prime\prime}\left(
1+\frac{3-\delta}{\beta E_{P}}\right)  }{\beta\left(  3-\delta\right)
}\right\}  .
\end{equation}
Always in the same limit, we can approximate the entropy as%
\begin{equation}
S\simeq\frac{2C_{r_{w}}E_{P}^{2}}{3r_{w}\beta}\pi^{2}\left[  \frac{\delta
}{6\left(  3-\delta\right)  }+\frac{2}{3}\right]  . \label{Sdelta}%
\end{equation}

\section{Acknowledgments}

The author would like to thank S. Liberati and G. Amelino-Camelia for useful
discussions and suggestions.

\end{document}